\documentclass[twocolumn,english]{IEEEtran}
\usepackage[T1]{fontenc}
\usepackage{color}
\usepackage{babel}
\usepackage{float}
\usepackage{amsthm}
\usepackage{amstext}
\usepackage{amssymb,graphicx,epstopdf}
\usepackage{caption}

\usepackage[unicode=true,
 bookmarks=true,bookmarksnumbered=true,bookmarksopen=true,bookmarksopenlevel=1,
 breaklinks=false,pdfborder={0 0 0},backref=false,colorlinks=false]
 {hyperref}
\hypersetup{pdftitle={Your Title},
 pdfauthor={Your Name},
 pdfpagelayout=OneColumn, pdfnewwindow=true, pdfstartview=XYZ, plainpages=false}

\makeatletter

\floatstyle{ruled}
\newfloat{algorithm}{tbp}{loa}
\providecommand{\algorithmname}{Algorithm}
\floatname{algorithm}{\protect\algorithmname}

\theoremstyle{plain}
\newtheorem{thm}{\protect\theoremname}

\usepackage{algorithm}
\usepackage{algorithmicx}
\usepackage{algpseudocode,balance}
\usepackage{subfigure}

\algnewcommand{\LineComment}[1]{\State \(\triangleright\) #1}

\makeatother

\providecommand{\theoremname}{Theorem}
\renewcommand{\baselinestretch}{0.985}
\begin{document}
\global\long\def\x{\mathbf{x}}
\global\long\def\pi{\theta_{i}}
\global\long\def\pj{\theta_{j}}

\title{Optimizing Average-Maximum TTR Trade-off\\
for Cognitive Radio Rendezvous}

\author{
Lin Chen$^{1,2}$, Shuyu Shi$^{1,3}$, Kaigui Bian$^2$, and Yusheng Ji$^3$\\
$^1$Department of Electrical Engineering, Yale University, New Haven, CT, USA\\
$^2$Institute of Network Computing and Information System, School of EECS, Peking University, Beijing, China\\
$^3$National Institute of Informatics, Tokyo, Japan\\
}

\maketitle
\begin{abstract}
In cognitive radio (CR) networks, ``TTR'', a.k.a. time-to-rendezvous, is one of the most important metrics for evaluating the performance of a channel hopping (CH) rendezvous protocol, and it characterizes the rendezvous delay when two CRs perform channel hopping. There exists a trade-off of optimizing the average or maximum TTR in the CH rendezvous protocol design. On one hand, the random CH protocol leads to the best ``average'' TTR without ensuring a finite ``maximum'' TTR (two CRs may never rendezvous in the worst case), or a high rendezvous diversity (multiple rendezvous channels).
On the other hand, many sequence-based CH protocols ensure a finite maximum TTR (upper bound of TTR) and a high rendezvous diversity, while they inevitably yield a larger average TTR. In this paper, we strike a balance in the average-maximum TTR trade-off for CR rendezvous by leveraging the advantages of both random and sequence-based CH protocols. Inspired by the neighbor discovery problem, we establish a design framework of creating a wake-up schedule whereby every CR follows the sequence-based (or random) CH protocol in the awake (or asleep) mode. Analytical and simulation results show that the hybrid CH protocols under this framework are able to achieve a greatly improved average TTR as well as a low upper-bound of TTR, without sacrificing the rendezvous diversity.
\end{abstract}


\section{Introduction}

``Rendezvous'' in \emph{cognitive radio} (CR) networks refers to the process for two secondary users (SUs) to find each other on a \emph{rendezvous/control channel} prior to data communications~\cite{akyildiz2006next}.
A \emph{rendezvous failure} occurs between two SUs when the rendezvous channel is unavailable due to the detection of primary users (PUs) or interference signals.

To alleviate the rendezvous failure problem, two types of \emph{channel hopping (CH)} rendezvous protocols have been widely used to create multiple rendezvous channels between
two SUs~\cite{bian2011asynchronous,chuang2013alternate,lin2011jump,zhang2011etch}. In a random CH protocol, two SUs hop across channels at random in search of each other. In a sequence-based CH protocol, each
SU starts a channel hopping process according to its own CH sequence
and local clock; two SUs' CH sequences are carefully chosen to spread
out rendezvous points over multiple pairwise common channels.
The \emph{time-to-rendezvous (TTR)} or rendezvous delay, is usually used for evaluating the performance of a CH rendezvous protocol.

There exists a trade-off of optimizing the average or maximum TTR in the design of a CH rendezvous protocol. The random CH protocol leads
to the best ``average'' TTR of $N$ timeslots given $N$ channels. However, two SUs may never rendezvous in the worst case, which implies that the
``maximum'' TTR can be infinite. Besides, the random CH protocol cannot guarantee multiple rendezvous channels, and the number of
rendezvous channels between two SUs is called the \emph{rendezvous diversity}. In contrast, many sequence-based CH protocols ensure a finite maximum TTR (upper bound of TTR) with a high rendezvous diversity, at the expense of incurring a large average TTR.

Naturally, we are particularly interested in the following question: \emph{Is it possible for an SU to determine when to switch to the random or sequence-based CH protocol for achieving the best performance in terms of both average and maximum TTR, while preserving a high rendezvous diversity?}

In wireless sensor networks, the wake-up schedule approach has been widely studied for addressing the neighbor discovery problem, which allows each
node to switch between two modes (awake or asleep) such that two neighboring nodes can maintain the link connectivity with energy constraints~\cite{bakht2012searchlight,dutta2008practical,kandhalu2010u}. Our research findings indicate that it is feasible to combine
the CH processes of random and sequence-based protocols by enforcing each SU to mimic the behavior of switching between two modes in a wake-up schedule.

In this paper, we strike a balance in the average-maximum TTR trade-off for CR rendezvous by leveraging the advantages of both random and sequence-based CH protocols. The contributions
of this work are summarized as follows.
\begin{enumerate}
\item We establish a design framework of creating a wake-up schedule whereby every SU follows the sequence-based (or random) CH protocol in the
awake (or asleep) mode, such that two SUs can achieve rendezvous with significantly improved average TTR, while an upper-bounded TTR and
rendezvous channel diversity are guaranteed as well.
\item We present a unified approach of devising a series of hybrid CH rendezvous protocols that  interleave random and sequence-based CH processes, and show that such protocols can achieve a high rendezvous diversity within an average TTR of $N$ slots, given $N$ channels.
\item Analytical and simulation results confirm that the hybrid protocols under the design framework are able to preserve a small average TTR as well as a low upper-bound of TTR, without sacrificing the rendezvous diversity.
\end{enumerate}


The rest of the paper is organized as follows. We provide the system model and formulate the problem in Section~\ref{sec:formulation}. In Section~\ref{sec:framework}, we describe the design framework of hybrid CH rendezvous protocols based on interleaving techniques. We evaluate our proposed framework using simulation results in Section~\ref{sec:simulation}, and conclude the paper in Section~\ref{sec:conclusion}.

\section{Problem Formulation}\label{sec:formulation}


\subsection{System Model}

We assume a CR network where each secondary user/node is equipped
with a CR operating over a set of orthogonal frequency channels that
are licensed to primary users. We denote each node in the network
by its unique identifer (ID), say $i\in\Lambda$, where $\Lambda$
is the set of all possible IDs ---i.e., the secondary node with its
ID $i$ is termed \emph{node $i$}. The set of channels which the nodes in this network can sense and operator over is called the \emph{sensible
channel set}, denoted by $C=\{1,2,3,\ldots,N\}$. The cardinality of the sensible
channel set is known as the \emph{sensible channel number},
or simply \emph{channel number}, written as $N=|C|$.

\begin{figure}[htbp]
\centering
\includegraphics[width=3in]{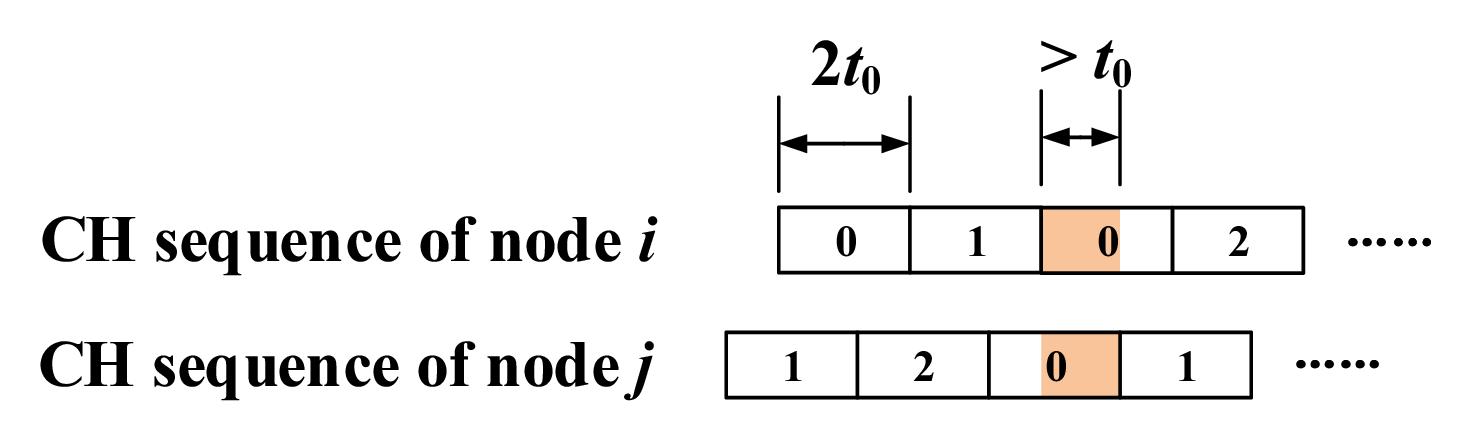}\caption{The CH sequences of nodes $i$ and $j$ are $\{0,1,0,2\ldots\}$ and $\{1,2,0,1\ldots\}$, respectively. They will rendezvous in the third slot on channel 0. Although the boundaries of their time slots are misaligned, since the duration of a slot is $2t_0$, the slot overlap on channel 0 is greater than $t_0$. \label{fig:align}}
\end{figure}

\textbf{CH sequence and clock drift}. We consider a time-slotted communication system
in which time is divided into consecutive time slots of equal length
$2t_0$, where $t_0$ is the time necessary for link establishment.
For example, as prescribed by IEEE 802.22 \cite{stevenson2009ieee},
$t_0=10\text{ ms}$ and thus time is divided into time slots of 20\ ms
in the IEEE 802.22 context. We double the link establishment time
to be the slot duration because two nodes can still have slot overlap
no less than $t_0$, which is adequate for link establishment, even
if slot boundaries are misaligned between them (please see Fig. \ref{fig:align})
\cite{gu2014fully}. Thus we can safely assume that slot boundaries
are aligned between two nodes.

\begin{figure}[htbp]
\centering
\includegraphics[width=3in]{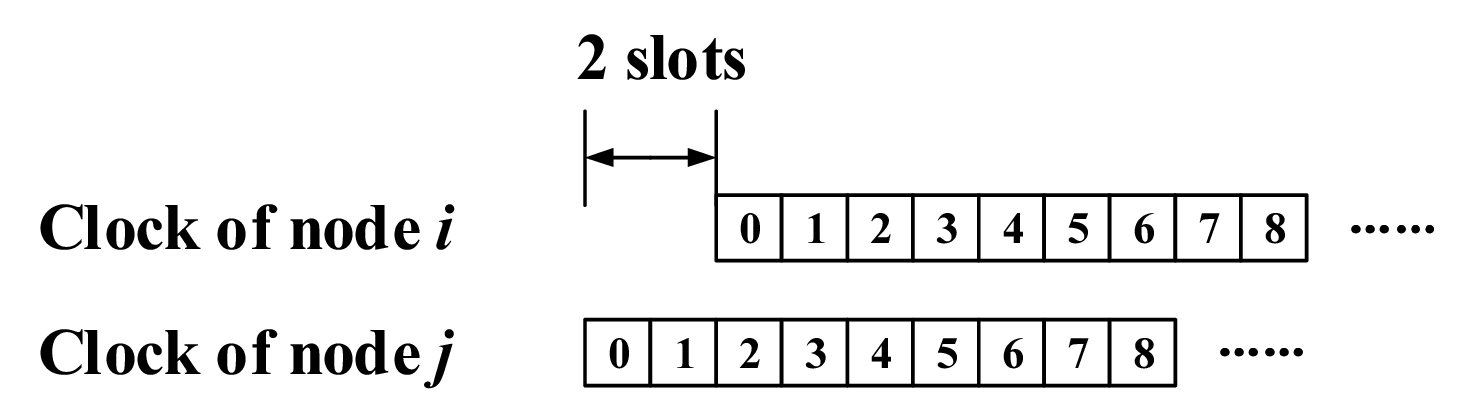}\caption{Nodes $i$ and $j$ have their respective local clocks. Time slots are divided in accordance with each node's local clock, numbered as slot 0, slot 1, slot 2,
etc. Local clocks of two nodes may differ from each other by a certain
amount of \emph{clock drift}---i.e., if node $i$'s clock is two slots
behind that of node $j$, then node $j$'s slot 2 will be its slot
0. \label{fig:clock_drift}}
\end{figure}

Each node has its local clock. Time slots are divided in accordance
with each node's local clock, numbered as slot 0, slot 1, slot 2,
etc. Local clocks of two nodes may differ from each other by a certain
amount of \emph{clock drift}---i.e., if node $i$'s clock is two slots
behind that of node $j$, then node $j$'s slot 2 will be its slot
0 (see Fig. \ref{fig:clock_drift}). In order to achieve rendezvous, each node
is capable of hopping across different channels in accordance with
its CH sequence and local clock. We formulate node
$i$'s CH sequence as a function
\begin{eqnarray*}
\theta_{i}:\mathbb{N}\cup\{0\} & \rightarrow & C\\
t & \mapsto & \theta_{i,t},
\end{eqnarray*}
where $\mathbb{N}\cup\{0\}$ denotes the set of time slots that begin
with slot 0 and $\theta_{i}$ maps each slot to a channel in $C$, which means that node
$i$ hops onto the frequency channel $\theta_{i,t}$ at slot
$t$.

\textbf{Rendezvous}. As aforementioned, there may
be a certain amount of clock drift between nodes $i$ and $j$'s local
clocks. We denote it by $\sigma_{ij}$, which means that node $i$'s
clock is $\sigma_{ij}$ slots behind that of node $j$. If $\sigma_{ij}<0$,
node $i$'s clock is ahead of node $j$'s in fact. If $\sigma_{ij}=0$,
they are synchronized. A rendezvous between nodes $i$ and $j$ is
said to occur if they hop onto the same frequency channel simultaneously.
Formally, a rendezvous is said to occur at node $i$'s slot $t$ if
$\theta_{i,t}=\theta_{j,t+\sigma_{ij}}$.
In this case, node $i$'s slot $t$ is called a \emph{rendezvous slot}
and the frequency channel $\theta_{i,t}$ is called a
\emph{rendezvous channel}.

Let $\mathcal{T}(\pi,\pj,\sigma_{ij})$ denote the set of rendezvous
slots given that node $i$'s clock is $\sigma_{ij}$ slots behind
that of node $j$. A noteworthy problem is whether the serial numbers
of rendezvous slots should be with regard to the local clock of node
$i$ or node $j$. In our definition, rendezvous slots are indexed
in accordance with the local clock left behind. For example, if node
$i$'s clock is behind that of node $j$ ($\sigma_{ij}>0$), then
rendezvous slots in $\mathcal{T}(\pi,\pj,\sigma_{ij})$ are with regard
to the clock of node $i$; otherwise, the clock of node $j$. If $\sigma_{ij}>0$,
$\{2,6\}\subseteq\mathcal{T}(\pi,\pj,\sigma_{ij})$ implies that nodes
$i$ and $j$ rendezvous at node $i$'s 2nd and 6th slots; if $\sigma_{ij}\leq0$,
it means that they rendezvous at the 2nd and 6th slots with respect
to node $j$'s clock. This idea is natural because the zeroth slot
of the clock left behind denotes when both nodes start channel hopping.

Given that node $i$'s clock is $\sigma_{ij}$ slots behind that of
node $j$, let $\mathcal{C}(\pi,\pj,\sigma_{ij})$ denote the set
of rendezvous channels between nodes $i$ and $j$.

\subsection{CH Protocol Design Problem}\label{sec:design-problem}

\textbf{CH protocol}.\textbf{ }A CH protocol is a fully distributed
algorithm whereby each node autonomously generates its CH sequence
only employing the information of its ID $i$. Formally, a CH protocol is a map
\begin{eqnarray*}
\theta:\Lambda & \rightarrow & \Theta\\
i & \mapsto & \pi,
\end{eqnarray*}
where $\Lambda$ is the set of all possible IDs and $\Theta=\{f:\mathbb{N}\cup\{0\}\rightarrow\mathbb{N}\}$
is the set of all CH sequences. Note that $\theta$ only depends on
the ID $i$ (i.e., $\theta$ is independent of the
clock drift between nodes, or the information of other nodes apart
from $i$.

Many existing CH protocols are \emph{periodic} in the sense that there exists a (minimum) positive constant integer $\tau$ (known as the period length) such
that \[\forall i\in \Lambda, t\in \mathbb{N}\cup \{0\}, \theta_{i,t+\tau}=\theta_{i,t}.\] The period length $\tau$ is a function of the channel
number $N$, denoted by $\tau(N)$. For example, CRSEQ \cite{shin2010channel} has a period length of $\tau(N)=N(3N-1)$ if $N$ is prime;
for the Jump-stay (JS) algorithm in \cite{lin2011jump}, the period length is $\tau(N)=3N$ if $N$ is prime.

\textbf{Performance metrics}. We introduce three performance metrics for evaluating the performance a CH rendezvous protocol.
\begin{enumerate}
\item \emph{Maximum time-to-rendezvous (MTTR)}. The latency before nodes
$i$ and $j$'s first rendezvous can be characterized by $\min\mathcal{T}(\pi,\pj,\sigma_{ij})$,
i.e., the minimal value in the set $\mathcal{T}(\pi,\pj,\sigma_{ij})$.
However, this latency relies on their clock drift $\sigma_{ij}$. In
practical scenarios, we are unaware of what the clock drift would
be, which, in fact, is a random variable. Thus we care about the worst-case
(maximum) latency given an arbitrary amount of clock drift, which
is exactly the \emph{maximum time-to-rendezvous (MTTR)}. Formally,
the MTTR between nodes $i$ and $j$ is given by
\[
MTTR(\pi,\pj)=\max_{\sigma_{ij}\in\mathbb{Z}}\min\mathcal{T}(\pi,\pj,\sigma_{ij}).
\]
MTTR is a bound for time-to-rendezvous. In the random channel hopping
protocol (RCH) \cite{bian2013maximizing}, each node $i$ randomly hops onto a channel
 at each time slot, and it is easy to show
that $MTTR=+\infty$ in this case---i.e., RCH fails to have a bounded
time-to-rendezvous or a guaranteed rendezvous diversity.
\item \emph{Average time-to-rendezvous (ATTR)}. Apart from MTTR, we take into account the average (expected) latency before the first rendezvous, termed as \emph{average time-to-time (ATTR)}. Formally, the ATTR between nodes $i$ and $j$
is given by
\[
ATTR(\pi,\pj)=\mathbb{E}[\min\mathcal{T}(\pi,\pj,\sigma_{ij})].
\]

\item \emph{Rendezvous channel diversity rate}. In practical applications,
some channels may encounter problems such as congestion, attack, eavesdropping etc.,
and thus become inappropriate for rendezvous and information exchange.
Ideally, we want to maximize the rendezvous diversity in hopes
that the two nodes can attempt to rendezvous on all channels. We introduce the metric called \emph{rendezvous
channel diversity rate}, which is defined for two nodes $i$ and $j$
such as
\[
div(\pi,\pj)=\min_{\sigma_{ij}\in\mathbb{Z}}\frac{|\mathcal{C}(\pi,\pj,\sigma_{ij})|}{N}
\]
and quantifies the minimum ratio of the number of rendezvous channels
to that of all sensible channels. It follows immediately
from the definition that the rendezvous channel diversity rate ranges
from 0 to 1.
\end{enumerate}

\textbf{CH protocol design problem}. The CH protocol design problem
is how to devise a fully distributed CH protocol $\theta$ whereby each
node $i$ autonomously generate its CH sequence $\pi$ such that the protocol
can achieve bounded MTTR, small ATTR and high rendezvous diversity
rate in CR networks
environments, in spite of a random clock drift between two nodes.

\section{Interleaving-based Framework for Hybrid CH Rendezvous Protocols}\label{sec:framework}

In this section, we propose a design framework for creating hybrid CH rendezvous protocols that minimize the ATTR of CH protocols while preserving their TTR bound.
We begin with introducing the neighbor discovery wake-up schedule design
problem. In accordance with a specified neighbor discovery wake-up
schedule, any existing CH protocol can be easily extended to a hybrid protocol under our framework.

\subsection{Definition of the Neighbor Discovery Wake-up Schedule}
In energy-constraint wireless sensor networks (WSNs), a neighbor discovery \emph{wake-up schedule} of node~$i$ is a binary sequence that consists of only zeros and ones,
\[
\x^{i}=\{\x_{0}^{i},\x_{1}^{i},\ldots,\x_{T_{i}-1}^{t}\}
\]
where $\forall0\leq t\leq T_i-1$, $\x_{t}^i\in\{0,1\}$. The length
of the sequence, i.e., $T_i$, is termed the \emph{period length} of
the wake-up schedule $\x^i$.  At the $t$-th time slot, if $\x_{t}^i=1$, the node will be active (awake); otherwise, it will be inactive (asleep).

The \emph{duty cycle} of a wake-up schedule quantifies the percentage of slots in which the node is active---i.e., the duty cycle of $\x^i$ is
\[
\frac{\sum_{t=0}^{T_i-1}\x_{t}^i}{T_i}.
\]
Obviously, the duty cycle ranges from 0 to 1. For instance, for $\x^i=\{0,0,1,1,0\}$,
its duty cycle will be $\frac{2}{5}=40\%$. For the purpose of energy saving, the duty cycle is supposed to be as small as possible.

We introduce the notion of \emph{cyclic
rotation} to characterize clock drift between nodes. For a wake-up
schedule $\x^i$, we define $rotate(\x^i,k)\triangleq \{\x'^i_{0},\x'^i_{1},\ldots,\x'^i_{T_i-1}\}$
where $\forall t\in[0, T_i-1]$, $\x'^i_{t}=\x_{(t+k)\bmod T_i}^i$.


Given any clock drift, a \emph{neighbor discovery} between nodes $i$ and $j$ successfully occurs if and only if $\forall k\in\mathbb{Z}$, $\exists t\geq0$ such that
\[
\x_{t\bmod T_{i}}^{i}=rotate(\x,k){}_{t\bmod T_{j}}^{j}=1.
\]

A \emph{neighbor discovery protocol}, $\nu$, assigns each node a wake-up schedule in accordance with its desired duty cycle in order to guarantee successful neighbor
discovery between any two neighboring nodes. If node $i$'s desired duty cycle is $\delta_{i}\in[0,1]$, then it will be assigned the wake-up schedule
$\x^{i}\triangleq\nu(\delta_{i})$. Formally, a neighbor discovery
protocol is a map $\nu$ from $\Delta\subseteq[0,1]$ to the set of
binary sequences, where $\Delta$, which is a subset of $[0,1]$,
is called the set of its \emph{supported duty cycles}, and $\nu$
is expected to satisfy that $\forall\delta_{i},\delta_{j}\in\Delta$,
$\forall k\in\mathbb{Z}$, $\exists t\geq0$ such that
\[
\nu(\delta_{i})_{t\bmod T_{i}}=rotate(\nu(\delta_{j}),k)_{t\bmod T_{j}}=1,
\]
where $T_{i}$ (or $T_j$) is the length of $\nu(\delta_{i})$ (or $\nu(\delta_j)$). Specifically,
$\forall\delta\in\Delta$, $\forall k\in\mathbb{Z}$, $\exists t\geq0$
such that
\[
\nu(\delta)_{t\bmod T}=rotate(\nu(\delta),k)_{t\bmod T}=1,
\]
where $T$ is the length of $\nu(\delta)$.

\subsection{Hybrid CH Protocols by Interleaving Random and Sequence-based CH Processes}

Suppose that $\nu$ is an arbitrarily given neighbor discovery protocol
and that $\theta$ is a periodic CH protocol with period length $\tau(N)$.

\textbf{Padding scheme}. According to the \emph{padding scheme}, we increase the channel number $N$ to some integer $N'\geq N$. We view the newly added $(N'-N)$
channels as aliases of the original $N$ channels. For example, if
the channel number is 3, we add a new channel, say, channel 4, so
that the new channel number amounts to 4. Channel 4 serves as a random
channel sampled from the original $N$ channels.

The first step of our proposed algorithm is to choose a supported
duty cycle, say, $\delta$, of the neighbor discovery protocol $\nu$.
Let $\x$ denote $\nu(\delta)$ and we write $T$ for the length of
$\x$. As aforementioned, $\x$ satisfies that $\forall k\in\mathbb{Z}$,
$\exists0\leq t\leq T-1$ such that
\[
\x_{t}=rotate(\x,k)_{t}=1.
\]
Then, by the padding scheme, the algorithm will slightly increase the channel
number $N$ to
$
N'=\min\{N'\in\mathbb{N}:N'\geq N,\gcd(\tau(N'),\sum_{t=0}^{T-1}\x_{t})=1\}.
$
Given $N'$ as the new channel number, with the CH protocol $\theta$
at hand, every node $i$ has its CH sequence $\theta_{i}$ with period
length $\tau=\tau(N')$, where $\gcd(\tau,\sum_{t=0}^{T-1}\x_{t})=1$.

\textbf{CH sequence generation}. We now present in Algorithm \ref{alg:CH-sequence-generating} how node $i$ generates its new CH sequence in accordance with
the wake-up schedule $\x$ and the original CH sequence $\theta_{i}$. Note that with the padding scheme, we can safely assume that the channel number $N$ satisfies that $\gcd(\tau,\sum_{t=0}^{T-1}\x_{t})=1$, where $\tau=\tau(N)$ is the period length of the CH protocol $\theta$---i.e., for convenience of notations, we simply use $N$ to denote the resulting slightly increased channel number (namely $N'$) after the padding scheme is conducted.

As demonstrated
in Algorithm \ref{alg:CH-sequence-generating}, at the $t$-th timeslot, the resulting new CH sequence generated by our proposed framework uses a slot that comes from the original CH sequence $\theta_{i}$ if the $(t\bmod T)$-th bit of $\x$ equals 1 (i.e., $\x_{t\bmod T}=1$); otherwise, it uses a random channel. We eventually obtain an interleaved new CH sequence $\theta_{i}'$.

\begin{figure}[htbp]
\centering
\includegraphics[width=3.5in]{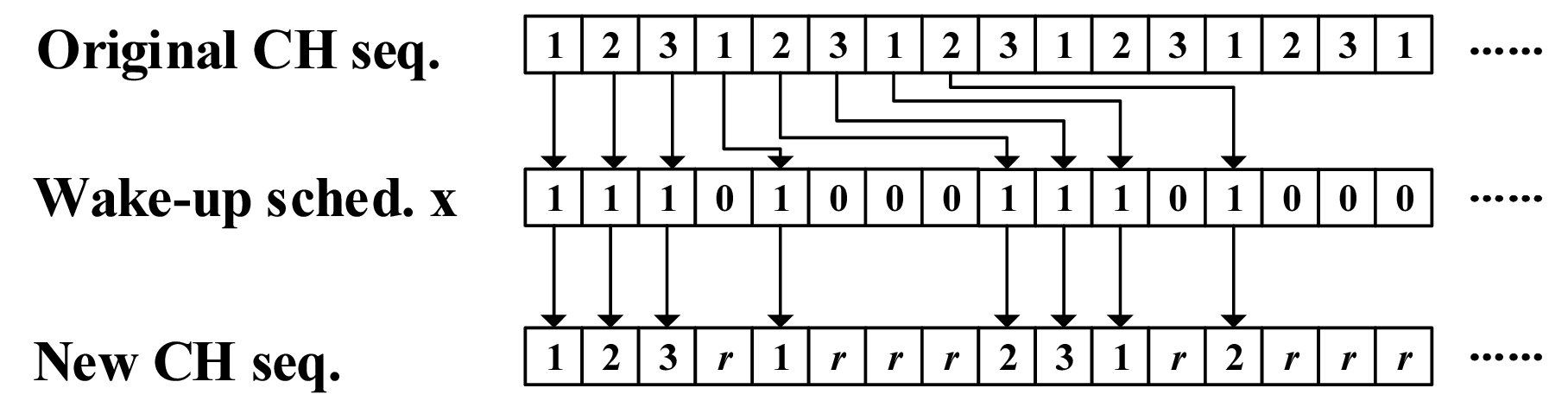}\caption{This figure illustrates an example of the CH sequence under the proposed framework. The original CH
sequence is $\{1,2,3,1,2,3,1,2,3,\ldots\}$ and the wake-up schedule is $\{1,1,1,0,1,0,0,0\}$.  The resulting new CH sequence is $\{1,2,3,r,1,r,r,r,2,3,1,r,2,r,r,r\ldots\}$, where ``$r$'' represents a randomly-selected channel. \label{fig:eg}}
\end{figure}

Fig. \ref{fig:eg} illustrates an example of the proposed framework. In the example, the original CH sequence is $\{1,2,3,1,2,3,1,2,3,\ldots\}$ and the specified wake-up schedule is $\{1,1,1,0,1,0,0,0\}$. We demonstrate the resulting new CH sequence generated in accordance with our proposed framework, where ``$r$'' represents a randomly selected channel.

The average TTR of random channel hopping is $N$. By interleaving
the random CH process, we can improve the average performance (i.e.,
ATTR) of the original CH protocol $\theta$. By leveraging the properties
of wake-up schedules, the proposed framework maintains a bounded TTR
and the rendezvous diversity inherited from the original CH protocol.

\begin{algorithm}[tbh]
\begin{algorithmic}[1]
\Require  Wake-up schedule, $\x$; original CH sequence, $\theta_i$.
\Ensure   New CH sequence, ${\theta}_i'$.
\State $T\gets |\x|$;
\LineComment{$T$ denotes the length of the wake-up schedule.}
\State  $t\gets 0$;
\LineComment{In the first place, initialize $t$ to $0$, which means that we begins generating the CH sequence from timeslot $0$.}
\State $t'\gets 0$;
\LineComment{$\theta_i$'s next slot that the algorithm is going to use is $\theta_{i,t'}$.}
\While{not rendezvous}
	\If{$\x_{t\bmod T}=0$}
		\State ${\theta}_{i,t}'\gets$ a random channel;
	\Else
		\State ${\theta}_{i,t}'\gets \theta_{i,t'}$;
		\State $t' \gets t'+1$;
        \LineComment{Update $t'$.}
	\EndIf
	\State Node $i$ hops onto channel ${\theta}_{i,t}'$ at the $t$-th time slot;
\EndWhile
\end{algorithmic}

\caption{CH sequence generating algorithm\label{alg:CH-sequence-generating}}

\end{algorithm}

Theorem \ref{thm:main} presents the main result regarding
the rendezvous performance improvement by the proposed method.

\begin{thm}
\label{thm:main}The new CH sequence generated by Algorithm \ref{alg:CH-sequence-generating}
has the following properties:
\begin{enumerate}
\item It can guarantee rendezvous within $\tau T$ slots (thus preserves
bounded TTR).
\item It achieves at least the same rendezvous diversity as the original CH protocol
(thus preserves the rendezvous diversity).
\item If the clock drift is $k$, let $B$ be $|\{t\in[0,T-1]\cap\mathbb{N}:\x_{t}=rotate(\x,k)_{t}=1\}|$,
then the ATTR of the new CH sequence, denoted by $ATTR_{\theta'}$
will be $\frac{B}{T}\cdot ATTR_{\theta}+(1-\frac{B}{T})\cdot N$
(thus when $\frac{B}{T}$ is small, $ATTR_{\theta'}\approx N$---i.e.,
it improves the ATTR).
\end{enumerate}
\end{thm}
\begin{IEEEproof}
Suppose that two arbitrarily given nodes, say, nodes $i$ and $j$,
have their original CH sequence $\theta_{i}$ and $\theta_{j}$ with
period length $\tau$. Without loss of generality, we assume that
the local clock of node $i$ is $k$ slots behind that of node $j$---i.e.,
the 0-th slot of node $i$ is the $k$-th slot of node $j$. Hereafter,
we number time slots in accordance with node $i$'s clock. Since $\x$
is a wake-up schedule, we have $\exists0\leq t_{0}\leq T-1$ such
that
\[
\x_{t_{0}}=rotate(\x,k)_{t_{0}}=1.
\]
We focus on the $(aT+t_{0})$-th slots, $a=0,1,2,\ldots$. Suppose
that in the $t_{0}$-th slot, node $i$ hops onto channel $\theta_{i,c_{1}}$
while node $j$ hops onto channel $\theta_{j,c_{2}}$. Let $A$ be
$\sum_{t=0}^{T-1}\x_{t}$. In view of Algorithm \ref{alg:CH-sequence-generating},
node $i$ hops onto channel $\theta_{i,c_{1}+aA}$ while node $j$
hops onto channel $\theta_{j,c_{2}+aA}$ in the $(aT+t_{0})$-th time
slot. Since $\gcd(\tau,A)=1$, we have
\[
\{((c_{1}+aA)\bmod\tau,(c_{2}+aA)\bmod \tau)\}_{a=0,1,2,\ldots,\tau-1}
\]
 is a permutation of
\[
\{(a,(c_{2}-c_{1})+a)\}_{a=0,1,2\ldots,\tau-1}.
\]
Consider
\[
\{(\theta_{i,a},\theta_{j,(c_{2}-c_{1})+a})\}_{a=0,1,2,\ldots,\tau-1}.
\]
Since $\theta$ is a CH protocol, given a clock drift of $(c_{2}-c_{1})$
slots, two nodes that use $\theta_{i}$ and $\theta_{j}$ as their
CH seequences respectively will rendezvous, say, on $\tilde{C}$ different
channels ($\tilde{C}$ is in fact the rendezvous diversity of the original CH protocol $\theta$) and $\tilde{T}$ different time slots in $\tau$ consecutive
slots. Therefore there are $\tilde{T}$ solutions to the equation
\[
\theta_{i,a}=\theta_{j,(c_{2}-c_{1})+a}
\]
w.r.t. $a=0,1,2,\ldots,\tau-1$. For $a=0,1,2,\ldots,\tau-1$, when
$\theta_{i,a}=\theta_{j,(c_{2}-c_{1})+a}$, $\theta_{i,a}$ can take
$\tilde{C}$ different values, which implies that the new hybrid CH protocol $\theta'$ under our proposed framework achieves at least the same rendezvous diversity as the original CH protocol.

Hence, we conclude that the new CH sequence can guarantee rendezvous
within $\tau T$ slots and that it preserves the rendezvous channel
diversity of the original CH sequence.

Now we calculate the ATTR of the new CH sequence. Let $B$ be
\[
|\{t\in[0,T-1]\cap\mathbb{N}:\x_{t}=rotate(\x,k)_{t}=1\}|.
\]
Suppose $X$ and $Y$ are uniformly random in $[1,N]\cap\mathbb{N}$
and independent, we have $\text{Pr}[X=c]=\text{Pr}[X=Y]=\frac{1}{N}$
for any fixed $c\in[1,N]\cap\mathbb{N}$. The ATTR of the new CH sequence
will be
\[
ATTR_{\theta'}=\frac{B}{T}\cdot ATTR_{\theta}+(1-\frac{B}{T})\cdot N.
\]
If $\frac{B}{T}$ is small, we have $ATTR_{\theta'}\approx N$.
\end{IEEEproof}

\section{Performance Evaluation}\label{sec:simulation}
\subsection{Simulation Setup}
In this section, we evaluate and compare the performance of existing protocols (e.g., CRSEQ \cite{shin2010channel} and
Jump-Stay (JS) \cite{lin2011jump}) and those hybrid protocols extended from our proposed framework (i.e., after the interleaving operation) via simulation results.

In our simulations, there are a total number of $N=11$ frequency channels and 20 pairs of nodes that need to rendezvous via channel hopping. Meanwhile, each node determines its clock time randomly and independently, yielding a stochastic clock drift between each pair of nodes. 
We simulated $X<N$ primary transmitters operating on $X$ randomly chosen channels. A
timeslot has a length of 10~ms. All secondary nodes are within the
transmission range of any primary transmitter.
In most existing work, it is assumed that a primary transmitter
follows a ``busy/idle'' transmission pattern on a licensed channel
\cite{geirhofer2008cognitive}, and we assume the same traffic pattern
here---i.e., the busy period has a fixed length of $b$ timeslots,
and the idle period follows an exponential distribution with a mean
of $l$ timeslots. A channel is viewed as ``unavailable'' when
PU signals are present in it. The PU traffic is the probability that PU signals are active in a channel under the specified traffic pattern. Once
two nodes hop onto a primary-user free channel in the same timeslot,
the rendezvous between them is established.


We conduct simulations while varying the duty cycle of wake-up schedules and the PU traffic. Specifically, we choose wake-up schedules with duty cycles being $5/14$, $7/14$, $9/14$, $13/14$ and $1$, under PU traffic of $25\%$ and $50\%$. Note that if the duty cycle of the wake-up schedule happens to be $1$, the resulting CH protocol is exactly the \emph{original CH protocol}---no random CH process interleaved. And we employ the three metrics defined in Sec. \ref{sec:design-problem} to evaluate the performance of rendezvous protocols.

\subsection{Maximum TTR}
The results for MTTR of original CRSEQ/Jump-stay and those resulting from our framework are shown in Fig.~\ref{figuremttr}. Note that if the duty cycle is $1$, the resulting CH protocol generated by our framework is exactly the original CH protocol (the original CRSEQ or Jump-stay).

We can observe that despite different PU traffic intensity and for both CRSEQ and Jump-stay, a larger duty cycle leads to an increased MTTR (i.e., worse performance). The original CRSEQ/Jump-stay has the longest MTTR compared with those generated by our framework.

This confirms that the proposed framework can preserve bounded TTR (shown in Theorem \ref{thm:main}) and improve MTTR in addition. With the import of random channel hopping slots, our framework achieves better worst-case performance.

\subsection{Average TTR}
We illustrate the results for ATTR of original CRSEQ/Jump-stay and those resulting from our framework in Fig.~\ref{figureattr}. It is noteworthy that a larger duty cycle positively correlates with ATTR for both protocols and under different PU intensity. Specifically, the original CRSEQ/Jump-stay (duty cycle equals $1$) has the worst performance in ATTR. This validates the theoretical analysis in Theorem \ref{thm:main}. A smaller duty cycle implies more stochastic ingredient (an increased number of random slots are interleaved into the resulting CH sequence), which, in turn, significantly improves the average performance. To summarize, the CH protocols generated by our framework outperform the original protocols (original CRSEQ/Jump-stay).
\subsection{Rendezvous Channel Diversity Rate}
\begin{figure*}[!htb]
\centering
\minipage{0.3\textwidth}
   \subfigure[Interleaving CRSEQ and random CH]{\includegraphics[width=2.2in]{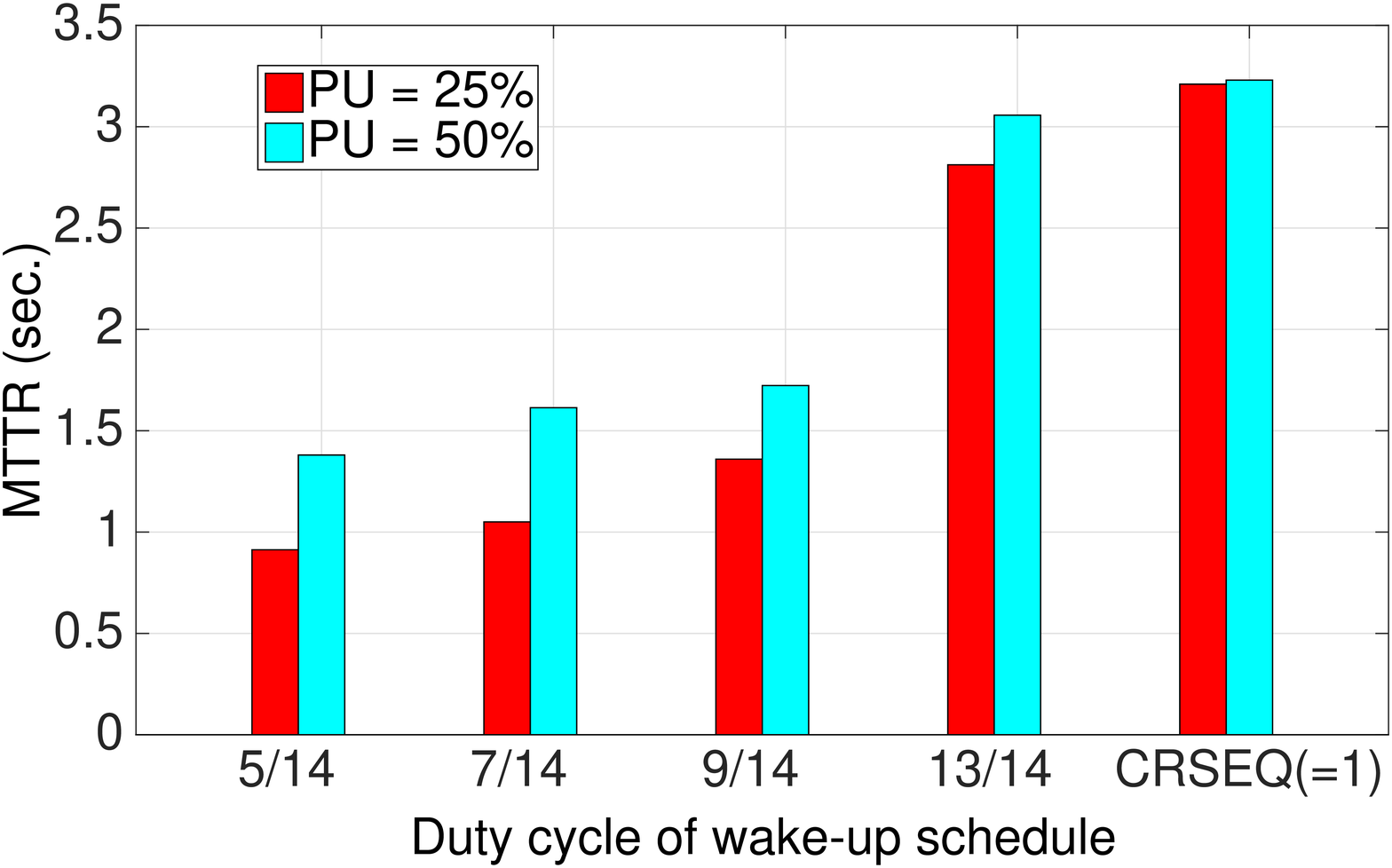}
     \label{figmttrCRSEQ}
     }
     \subfigure[Interleaving Jump-stay and random CH]{\includegraphics[width=2.2in]{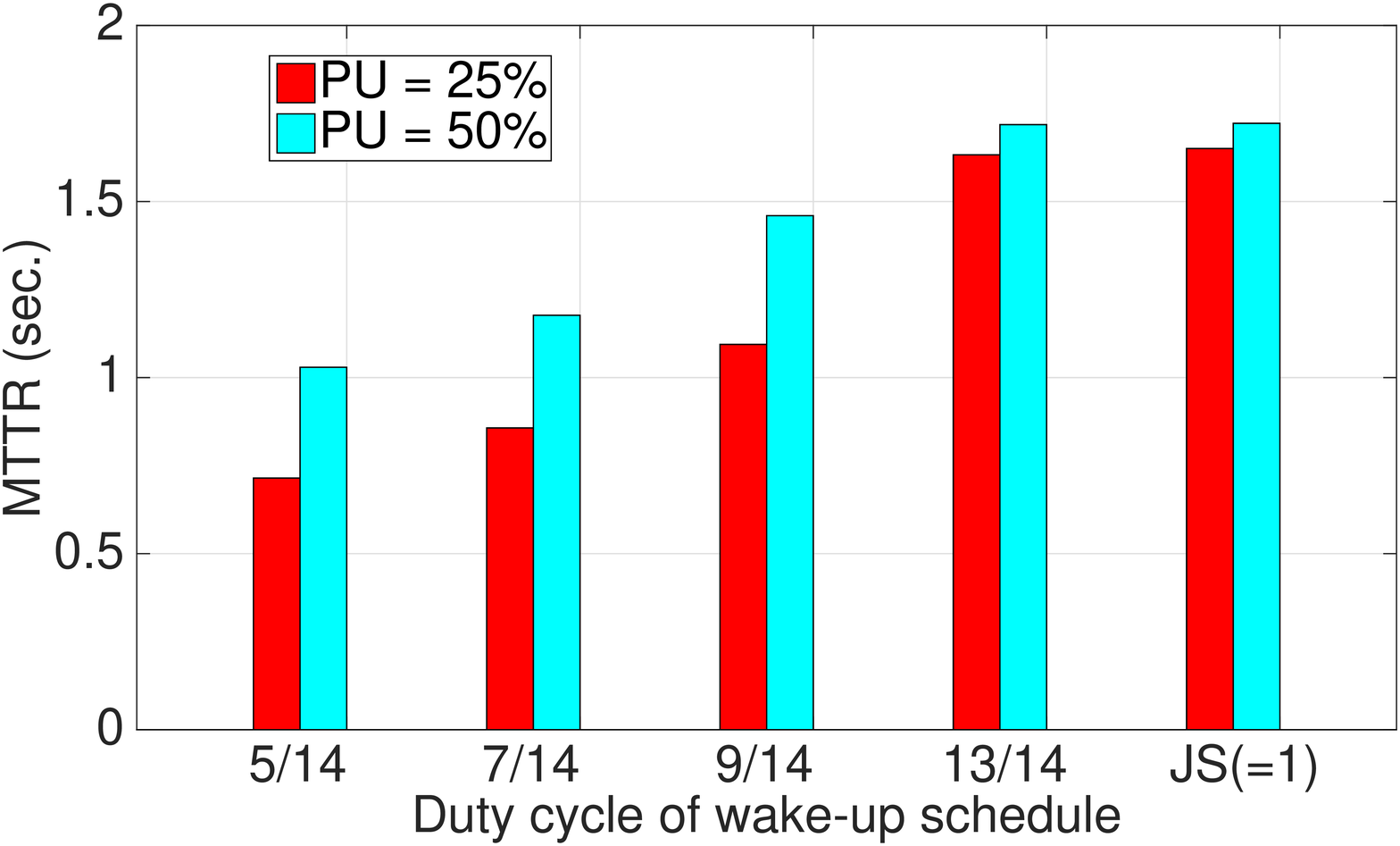}
     \label{figmttrjumpstay}
     }
   \caption{Maximum TTR of hybrid protocols (duty cycle smaller than 1) and the original CRSEQ/Jump-stay protocol (duty cycle equals 1). We interleave CRSEQ (or Jump-stay) with random channel hopping in Fig. \ref{figmttrCRSEQ} (or Fig. \ref{figmttrjumpstay}).}\label{figuremttr}

\endminipage\hfill
\minipage{0.3\textwidth}
 \subfigure[Interleaving CRSEQ and random CH]{\includegraphics[width=2.2in]{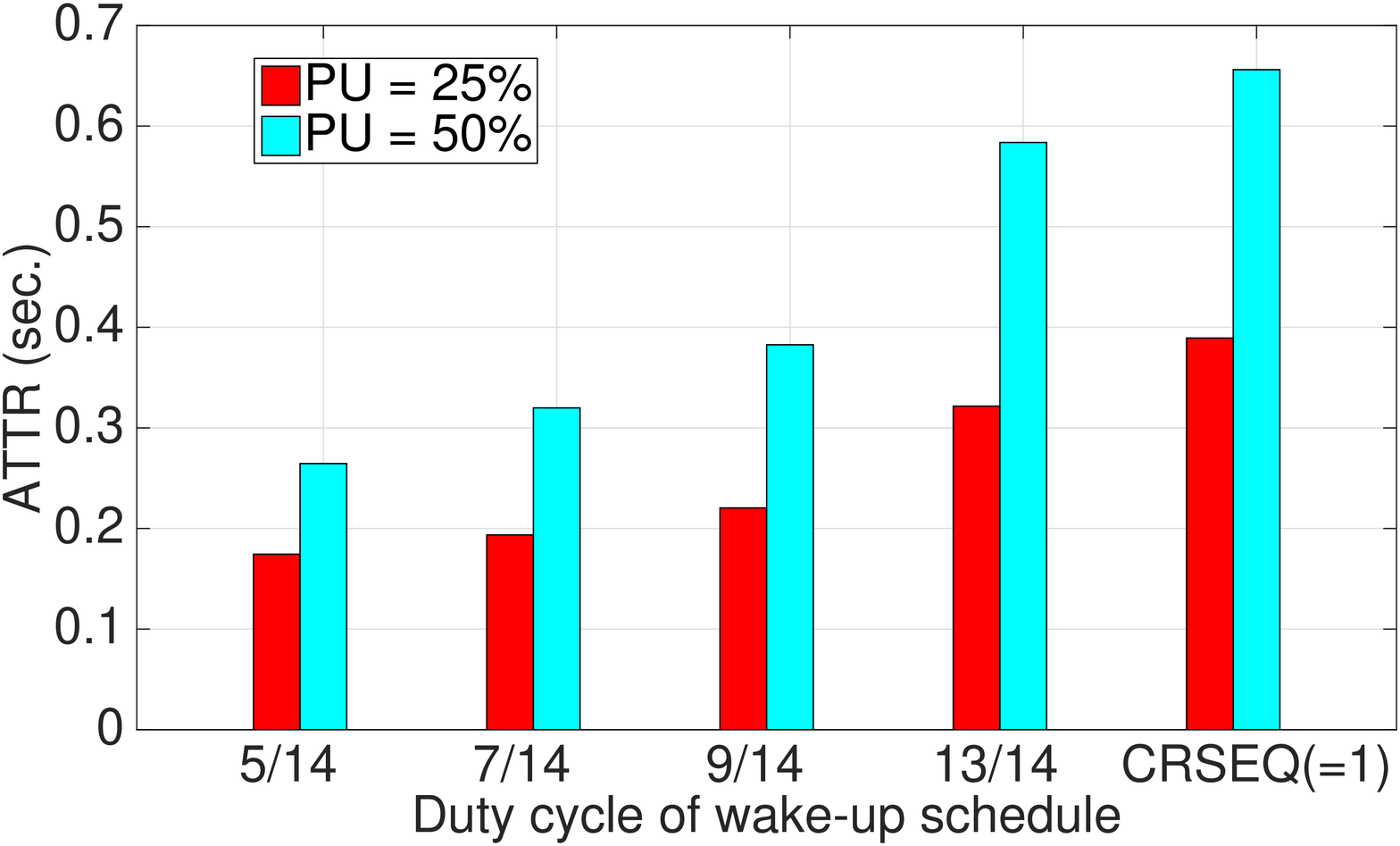}
     \label{figattrCRSEQ}
     }
     \subfigure[Interleaving Jump-stay and random CH]{\includegraphics[width=2.2in]{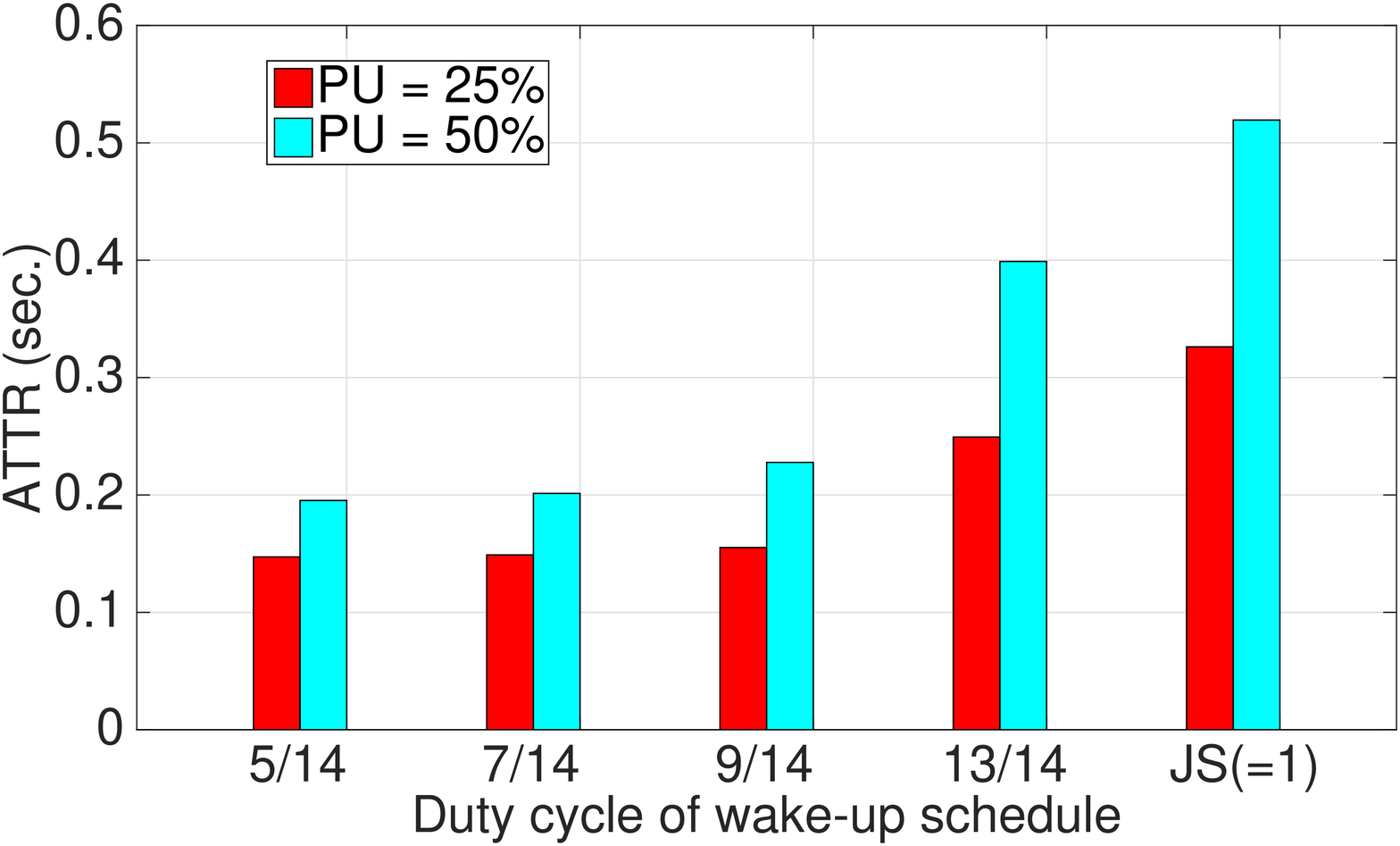}
     \label{figattrjumpstay}
     }
   \caption{Average TTR of hybrid protocols (duty cycle smaller than 1) and the original CRSEQ/Jump-stay protocol (duty cycle equals 1). We interleave CRSEQ (or Jump-stay) with random channel hopping in Fig. \ref{figattrCRSEQ} (or Fig. \ref{figattrjumpstay}).}\label{figureattr}
\endminipage\hfill
\minipage{0.3\textwidth}
 \subfigure[Interleaving CRSEQ and random CH]{\includegraphics[width=2.2in]{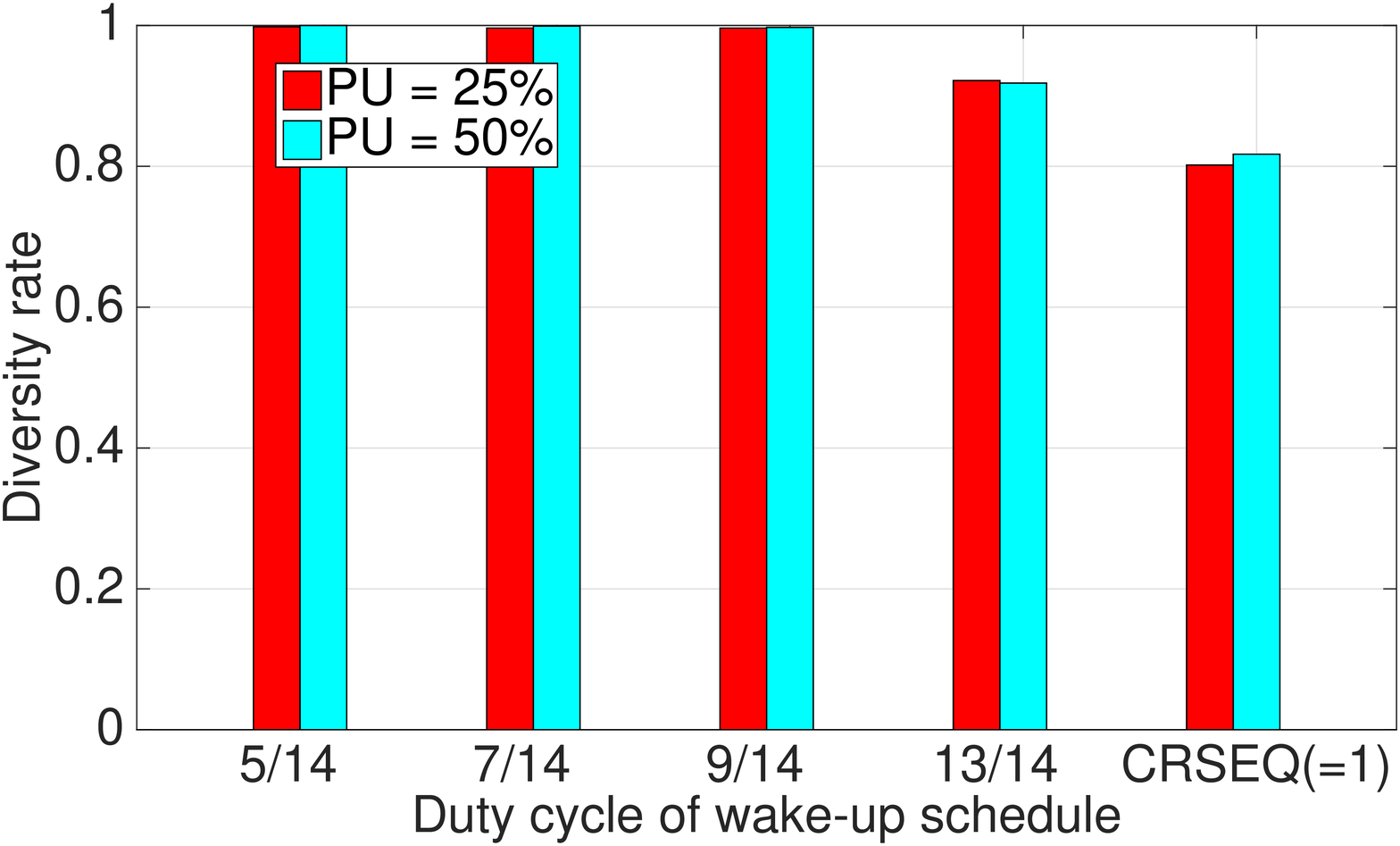}
     \label{figdivCRSEQ}
     }
     \subfigure[Interleaving Jump-stay and random CH]{\includegraphics[width=2.2in]{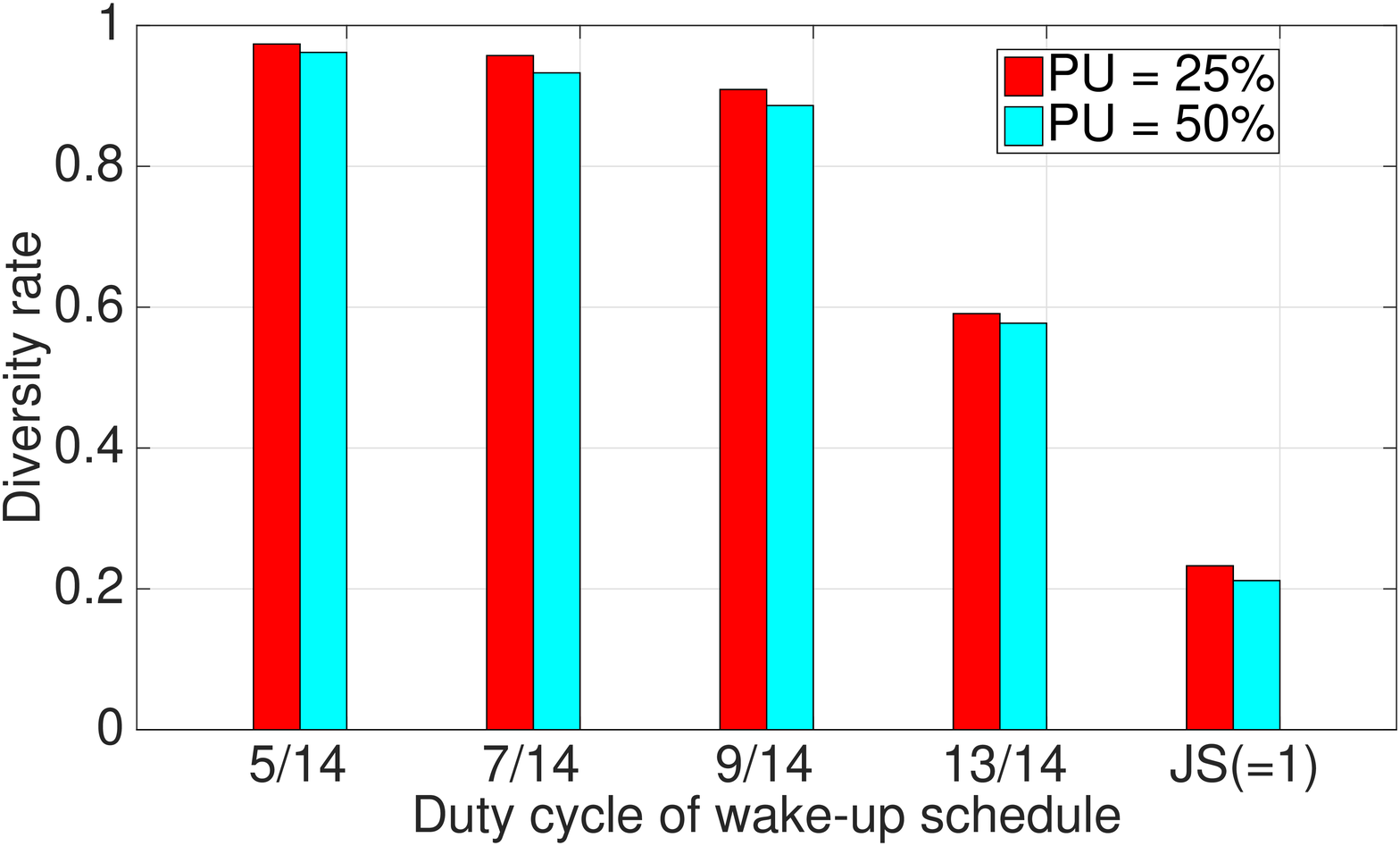}
     \label{figdivjumpstay}
     }
   \caption{Diversity rate of hybrid protocols (duty cycle smaller than 1) and the original CRSEQ/Jump-stay protocol (duty cycle equals 1). We interleave CRSEQ (or Jump-stay) with random channel hopping in Fig. \ref{figdivCRSEQ} (or Fig. \ref{figdivjumpstay}).}\label{figurediv}
\endminipage
\end{figure*}

Rendezvous channel diversity rate is a metric for measuring a CH protocol's ability to establish rendezvous in varied channels. The results for diversity rates of original CRSEQ/Jump-stay and those resulting from our framework are presented in Fig.~\ref{figdivCRSEQ} and ~\ref{figdivjumpstay}, respectively. It can be observed that diversity rate declines as the duty cycle increases and that the original CRSEQ/Jump-stay has the worst/smallest diversity rate. This observation supports the conclusion of Theorem~\ref{thm:main} that our proposed framework can preserve diversity rate. In addition, it validates that by introducing stochastic CH slots, we can obtain an improved diversity rate.

\section{Conclusion}\label{sec:conclusion}
In this paper, by leveraging the properties of neighbor discovery wake-up schedules, we establish a design framework for creating a series of hybrid CH protocols whereby every SU employs the specified wake-up schedule to interleave the sequence-based (or random)
CH protocol in the awake (or asleep) mode. Analytical and simulation results show that the hybrid CH rendezvous protocols under the proposed framework can significantly improve average time-to-rendezvous and preserve a low upper-bound of TTR and the rendezvous channel diversity simultaneously. It is also validated by extensive simulation results that our method remarkably outperforms existing CH protocols, CRSEQ and Jump-stay.
\bibliographystyle{abbrv}
\balance
\bibliography{mycollection}

\end{document}